# OntologyNavigator: WEB 2.0 scalable ontology based CLIR portal to IT scientific corpus for researchers.

### Gérald Kembellec

Paragraphe Laboratory
Paris 8 University.
2 rue de la Liberté
92526 Saint-Denis Cedex, France
gerald.kembellec@univ-paris8.fr

### Imad Saleh

Paragraphe Laboratory
Paris 8 University.
2 rue de la Liberté
92526 Saint-Denis Cedex, France
imad.saleh@univ-paris8.fr

### Catherine Sauvaget

LIASD Laboratory
Paris 8 University.
92526 Saint-Denis Cedex, France
cath@ai.univ-paris8.f

**Abstract**

This work presents the architecture used in the ongoing OntologyNavigator project. It is a research tool to help advanced learners to find adapted IT papers to create scientific bibliographies. The purpose is the use of an IT representation as educational research software for researchers. We use an ontology based on the ACM's Computing Classification System in order to find scientific articles directly related to the new researcher's domain without any formal request. An ontology translation in French is automatically proposed and can be based on Web 2.0 enhanced by a community of users. A visualization and navigation model is proposed to make it more accessible and examples are given to show the interface of the tool. This model offers the possibility of cross language query. Users deeply interact with the translation by providing alternative translation of the node label. Customers also enrich the ontology node labels with implicit descriptors.

## 1. Introduction

The purpose of this article is to develop a method for approaching the IT field for the use of student researchers, typically in their second year of master or beginning their PhD thesis. In the French university context, it is common to see students in $2^{nd}$ and $3^{rd}$ cycles experiencing real difficulties in collecting the documentation on their field of study or research. The main idea is to support these learners by a tool to supplement their perception of the knowledge domain "stored" in ontology. It is to consider, even hoping, that mastering tool lets it become obsolete for the advanced learner (became autonomous). The context is bibliographical research in the IT area for "Insiders" but not experts, who are increasingly lost in a predominantly English corpus. Once the subject and the target audience have been defined, it is necessary to determine the habits, practices, behaviour and attitudes of young researchers in their research of information. An important point of transition between the first and the second year of master is the increased focus on access to quality and credit worthy information. The purpose of information retrieval systems, such as conventional search engines or documentaries on the Internet is to search by keywords or natural language. During the first part of their studies, students tend to seek information on the total mass of the Internet without particular method, discernment or qualitative discrimination. When transitioning to a higher level of studies the method of research evolves. This is mainly due to the fact that access to a huge amount of information causes two major problems. The first major obstacle to efficient research is the difficulty to control the quality of information. The ability to assess the creditworthiness of the information will depend on the level of knowledge of the student researcher. The second problem identified in a classic request on a search engine is the amount of information returned by each search. While not purporting to replace the role of a director of research, OntologyNavigator seeks to lead the researcher in his approach of the field of IT research. In this article, the concept of domain ontology is defined. The way that domain ontologies can help to search scientific information is shown. This paper describes step by step the design of the tool, how OntologyNavigator includes translation tools, and discusses the methods of representation. The article continues with examples of the use of the tool.

## 2. DOMAIN ONTOLOGIES AND INFORMATION RETRIVAL

In the context, domain ontology means a conceptual hierarchy designed by an expert within a structure, where elements are linked by their proximity in terms of syntactic or semantic relations. The traditional approach to the use of ontology's area is to prioritize subsets of the area for management purposes. The ontology is then used most often to prioritize and rank the components of the domain and to describe their relationship. A frequent application is indexing a specialized corpus. A more innovative use of ontology is to reverse the process. It is possible to use domain ontology as a means of research in a text, a corpus, a digital library, or perhaps even the Internet. Thanks to a combination of different semantic technologies, Stephan Bloehdorn has proposed an interesting method of searching in digital libraries (Bloehdorn, S. et al., 2007). He defines an approach by analysis of structured questions in natural language with a formal grammar. It is then the role of the system to understand the question, identify keywords, titles and authors. Basic examples of questions would be: who wrote this book? What book deals with this specific topic? Which article is part of this conference and corresponds to these keywords? This approach translates natural language into meta-data, and rephrases the question in SPARQL language (W3C, 2008, SPARQL). The answer is contained in a Resource Description Framework (RDF) file (W3C, 2004. RDF/XML Syntax Specification), this enables updates in real time, the use of a variety of formats as well as sourcing from many locations. This method enables the user to avoid using any database in the common sense of the term.

IT is a very broad area, which includes a multitude of sub-disciplines, and is a powerful tool used in many scientific fields. It is therefore necessary to understand the nature and context of the user's research and his angle of research as much as possible. For example: the couple of words "data storage" will not have the same meaning for an assembly technician, a systems and networks engineer or a librarian. The technician's perception of "data storage" is the hard disk or USB drive. The systems and networks engineer will have a broader vision of "data storage", not only the concept of devices, but also methods of storage such as NAS, data redundancy (RAID level), information sharing techniques (NetBIOS, NFS, SMB ...) and permission(s) (reading, writing and execution). Finally, the librarian will understand the term "data storage" primarily as an integrated library system, which administers the loans and reservations and manages the order tracking and state of the inventory. These three professionals, having advanced knowledge in their own particular field have different uses of the term "data storage". This is however not a case of polysemy (multiple meanings) but is rather a difference in the angle of perception of these three professionals. The issue of user relevance arises in the particular case of the IRS. The idea of user relevance has greatly influenced the tool, which focuses on the angle of perception of the user and not only on the data.

### 3 TAXONOMY BASED IT ONTOLOGY

Initially, the approach will be onomasiological or top-down, i.e. the corpus will be classified in a structure, which is a finished standardized set. In a second step, the structure is enriched, where this becomes necessary, through the ad junction of additional corpus. The domain ontology consists of a tree of topics ranging from a generic root (in this case computer science) to the leaves of knowledge. The arcs are links between nodes that materialize top-down or bottom-up relations or ties of similarity. The ontology contains no articles, but nodes with labels containing keywords issued by superior nodes. These keywords can generate a request to be submitted to the main scientific on-line libraries.

#### 3.1 Proposal of a model

This project must consist in a tool of flexible use, which integrates the field of a particular user to be within the user's grasp. Therefore it should help the user mastering his field of expertise. This tree can simply be seen as the external skeleton or exoskeleton in the IT field. First keywords of each node or leaf are the words constituting its label. These keywords are called "native" keywords, as opposed to other keywords added afterwards, which will be referred to as "added" keywords. The starting point was the description of research with a minimal ontological exoskeleton. To put into place such a minimal ontological exoskeleton it was necessary to find taxonomic approaches representing as carefully and as fully as possible the broad field of IT. Then, to conceptualize this field, it was necessary to segment the titles of each branch. This specification phase passes through a stage of construction of keyword "clusters" related to each branch, thanks to lemmas (canonical form of a lexeme) extracted from titles. From a technical point of view, for greater ease of handling, it would be appropriate to integrate the ontology and its keywords in a database, which will result in a comprehensive ontology in Extensible Mark-up Language (W3C, 2008, Canonical XML) where developments are updated in real time. For the test phase, the corpus of research will be composed of the titles of articles published since 1945 and referenced in the Database systems and Logic Programming (Ley, M. & Reuther, P., 2006) by Michael Ley from the German University of Trier. It is the source of an XML document of about one million admissions in BibTeX[1] format (format of bibliographic description of L$^A$TEX[2]). It should be noted that the papers are written in various languages (3.6). It also proposes meta-queries to on-line digital libraries such as Computer Science Bibliography[3] or ACM.

---

[1]     http://www.bibtex.org/
[2]     http://www.latex-project.org
[3]     http://liinwww.ira.uka.de/bibliography

### 3.2 Choosing the best reference for IT classification

From a technical point of view, to increase handling comfort, the ontology and its keywords were integrated in a database. That results in a comprehensive XML ontology where developments are updated in real time. The model tried as a first step to find an agency specialized in computer sciences. Then it proposed a system of representation in the field that the tool wishes to model. For the sake of simplicity lit take the on-line encyclopaedia Wikipedia as a first step. Indeed Wikipedia from an IT perspective is classified according to an internal hierarchy, has an abundant corpus and is immediately available in XML and RDF. Unfortunately, as of today the scientific legitimacy of Wikipedia is not demonstrable. Let us then turn to Computing Classification System (Association for Computing Machinery, 1998), whose legitimacy is evident. Moreover, conveniently the Association for Computing Machinery (ACM) has its own digital library of scientific articles indexed according to the CCS model. However, the CCS is not usable as it stands. The CCS is more in the state of taxonomy than ontology. According to Grüber, an important aspect of ontology (in addition to clarity, consistency, minimal commitment, and deformation) is scalability (Grüber, T. R., 1995).

### 3.3 A WEB 2.0 way to enrich the ontology with keywords and implicit descriptors

Consider the corpus as a mass of papers' titles in terms of information and statistical science. According to Le Coadic (Le Coadic, Y.F., 2005), when a serie of scientific articles is considered, a specific attention to meaningful words and their co-occurrence must be given in order to generate significant semantic proximities. So when a couple of associated words emerge simultaneously in several node labels, it is likely that the subjects in question are associated. Of course, in this case this approach is only used on titles. Nevertheless, ACM labels appear sufficiently precise to be representative of all documents, both from the general and particular point of view. Thus, the words that best represent the label will be added as keywords to get to the document and to the branch of the ontology. Other words, less representative, will be added as "semantically near". Subsequently, during the indexing phase of a digital library, if an article appears to be indexed in two places, it is proposed to establish proximity link between the two branches.

Dislike Stephanov's work for building IT domain ontology with ACM CCS (Stefanov K. & Kornelia T. 2003) who deleted all "miscellaneous" and "general" nodes, this approach allows greater compatibility with the tools offered by ACM. Indeed, the implicit subject descriptors sometimes refer to these nodes. In addition, the portal uses the ACM full classification described for his papers. A "general" node is used (at any level) if the article covers most of the concepts in an area. If a paper cannot be classified under any other node, then the "Miscellaneous" node of the given area is used (Association for Computing Machinery 1998).

The lexical scope of the original taxonomy has been expanded by subject descriptors, which only specify certain nodes. This document is available in text format on the ACM website. Implicit Subject Descriptors (also called "Proper Noun Subject Descriptors") are proprietary names of products, systems, languages, and prominent people in the computing field, along with the category code under which they are classified. For example, "C++" is under "D.3.2 Language Classifications" (Association for Computing Machinery 1998).

This list of descriptors was translated into XML for the purpose of the tool. These descriptors are integrated in the ontology as "implicitDescriptorOf" type leaf specifying certain nodes (cf. Figure 1). Sometimes there are co-occurrences of descriptors, which automatically create semantics arcs between nearby nodes described. Besides the fact that these specifications consistently improve research, they also create new semantic arcs integrated to the ontology. This theoretical approach, although functional, is inadequate to enrich the ontology while remaining close to the user. Indeed, this tool is designed to adapt to the user by exchanging knowledge. But in the case previously described, this specification is achieved by the system for the final user. The second use of these descriptors is to provide the users with the opportunity to submit their own interpretation. These synonyms are then classified as keywords and stored for future researches. These keywords will not be able to be seen during browsing, but they will be included for the future researches. The process will also be applied when added co-occurrences expressions. Creations of "IsRelatedTo" arcs between keywords and descriptors will be created.

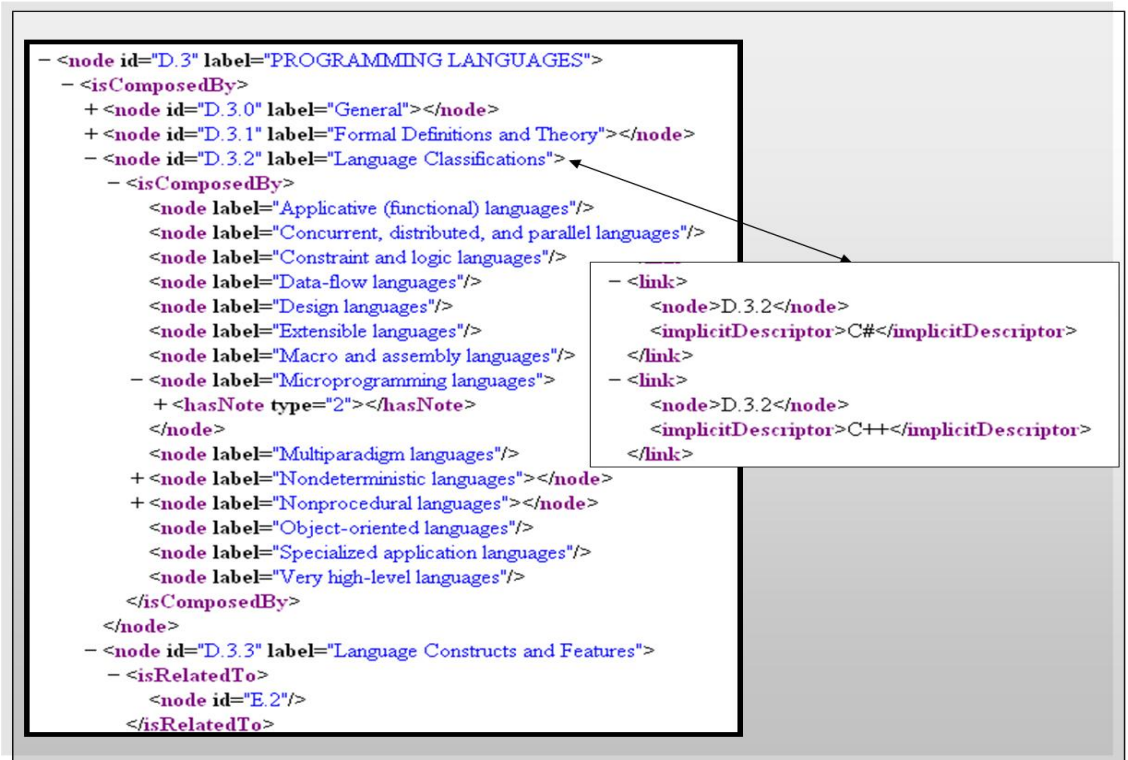

Figure 1. ACM CCS short sample and an XML translation of the native implicit descriptors list

It is important to notice that implicit descriptors are different of generic keywords of the CCS that can be found on ACM portal. On the one hand, generic keywords are explicit descriptors of CCS nodes. They precisely describe the main concept of a node. On the other hand, implicit descriptors specify very deeply a concept with a name of a major actor or the main software in a field of knowledge.

### 3.4 Corpus Interface

In the long run, a method of compilation (or clustering) from different on-line databases of articles like CSBIB, DBLP, ACM and others will be done. The clustering will go through a phase of pre-treatment. Each library has its own scientific query interface; we will try to find the RDF document that describes each database. It should be noted that if each site provided data Description services such as RDFa (W3C, 2008, RDFa), this work would be greatly simplified. A database, the "The scientific library of the field of Information Technologies" will be created, describing each article by its title, the context and year of publication, and authors of this article. The database, automatically updated each week on an incremental basis, would ideally continuously generate a single RDF document describing the pseudo corpus. The term "continuously" means that in theory for every query, a snapshot of the corpus will be established by RDF through interrogation of the database and will be processed to reflect the weekly updates. This approach, while desirable is technically unrealistic. It is possible and even desirable in view of the vast amount of data (in a resource-saving system) to maintain a snapshot cache. This snapshot of the database would become a RDF file and therefore the representation of the pseudo corpus. The corpus of scientific articles would not be hosted locally on the host machine of the ontology for legal reasons, but also for reasons of storage capacity. This is why the term pseudo-corpus is used rather than corpus. Indeed labels, and possibly abstracts indexed in digital libraries do not strictly constitute a corpus.

### 3.5 The perspectives for IT ontology

The index of the pseudo corpus is composed by titles of articles. Most of the times the titles of scientific articles are long enough to provide a number of keywords indicating the leading ideas. During the phase of indexing the corpus, if an article's title appears as "unclassifiable", we propose to classify it momentarily in a branch of the ontology having the closest semantic proximity within a "miscellaneous" or "general" subsection. Then once the ontology has reached a sufficient size, the "orphan" article will be classified permanently by creating a new branch on the ontology where semantic proximity is the greatest (using added keywords). The process described above is one of the vectors of the evolution of an ontology, which is not static but evolves with the corpus and the work of the users and experts. The extensions that may be added to the ontology must be anticipated during its creation. It should be possible to add new concepts without having to modify the foundation of ontology. For example a "branch" which turns up an important number of common keywords would constitute a suitable root for the ontology. It may be possible to automate this task e.g. if a text generative algorithm would be able to produce a full text leaf of the ontology. This algorithm should be able to redistribute keywords extracted from one or more articles classified in a temporary general branch of the ontology.

### 3.6 WEB 2.0 Cross Language module implementation

The *lingua franca* of scientific research today is English. Each researcher should in theory feel comfortable with this international scientific language. Why is it important to make an effort to translate the titles of the IT ontology branches in vernacular language (French in the context) while the corpus is mainly English, the predominant scientific language? However, even if the user feels comfortable reading technical and scientific texts (as the case may be with a good dictionary in hand), he may feel more at ease in French to conduct his research.

In order to create a more customizable tool, a Web 2.0 approach was used, i.e. hybrid translation starting with automatic translation, which is thereafter corrected and completed through communal manual translation. The simplest and most economical solution to automate an Anglo-French translation would be to use an on-line translation tool. The more used tools are Babel fish, Yahoo and Google Translate. An API was written to generate a French version of the ontology based on one of these tools. It can be pointed out that this kind of on-line applications would benefit from having its own official API. Of course nothing can replace manual translation, which is why a notion of folksonomy was incorporated with a RDF Site Summary (World Wide Web Consortium, 2002, RSS 2.0 specification) in the tool. This enables the last user to report a translation error, or imprecision, to the management committee. This group will consist of researchers from laboratories of the research and training unit, which will validate the proposal or reject it. According to Thomas Vander Wal, the value of external marking of the folksonomy comes from the users using their own words, which add an explicit dimension, which will be an inference of the object (Vander Wal, T., 2006). The system of translation of the ontology's nodes automated in a first step continued and developed by English language users and validated by experts, can be carried out without recourse to a professional translator, or occupying an expert on a full-time basis. This procedure implies considerable time saving for researchers and the financial economy should not be underestimated. The technical aspects of this process should be simplified, as much as possible, for the user so as not to discourage him from making a proposal e.g. making a proposal should also not take him more than a few seconds. Once the proposal has been submitted (cf. Figure 2), an RSS feed is generated and will remain active until at least two committee members have verified the proposal. It is contemplated to correct the French part of the ontology over a period of time which is yet to be determined. Another advantage of this process is that it takes into account terminology modification, which is inherent to the field of IT. Due to the interaction between the system and the user, the user enriches his knowledge in the field in question while participating in its evolution.

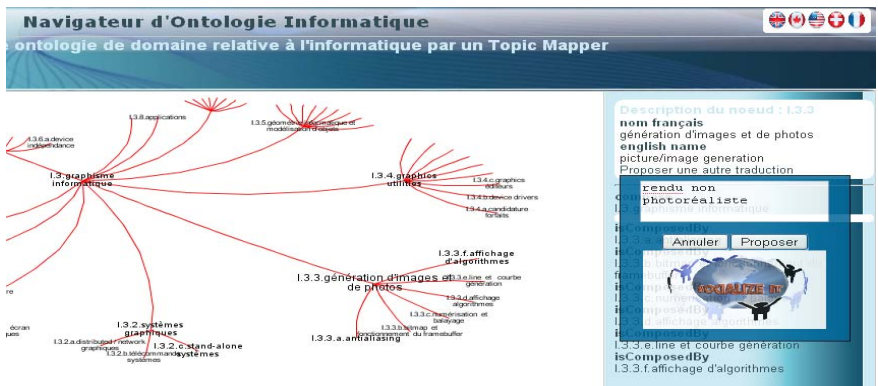

Figure 2. Alternative translation of a node

### 3.7 Choosing a model of representation

To make the corpus more accessible, to facilitate the representation, the domain ontology was represented as a navigable map. The tree should enable the user to focus on the branch containing a formalization of the concept sought. There are many ways to view ontologies, but all are not specific to navigation, at least not as concerns intuitive navigation. In this context, the tool of representation must abide by a number of rules set out by Christophe Tricot and Christophe Roche (Tricot, C. & Roche, C., 2006). To be effective a visualization system should observe the following rules as a minimal requirement:

- Provision of an overview of the ontology. This will allow the user to identify all the concepts in the field.

- Use of a "focus + context" to allow the user to concentrate on certain aspects while having access to others;

- Use of plane geometry, to avoid disturbing natural perception. This particular point, has however not been followed in the present case, because giving the mass of data to display and the wish to comply with the other principles, it is complex, if not impossible, to combine a tree display and Euclidean geometry.

According to feedback C. Tricot obtained from an experiment, two types of users emerge: "newcomers" and "experts". Newcomers understand the field and its concepts without perceiving details of the organization and interactions. Experts have a perfect mastery of the entire field both in terms of the content of the concepts and the links that bind them together. For the target audience, users have a profile of a master student or a PhD beginner who searches scientific information on a subject in a specialized field. A compromise on the representation of the field was found. It offers a direct access to context on the element in focus. In the article by C. Tricot, it appears that the model representation by radial tree is the most suitable for experts and the model representation by eye tree is the most suitable for newcomers. The eye tree visualization allows a global view of the field and the possibility of a wide-angle focused (fish eye polar) on a point of detail around which the field is articulated. The main shortcoming of this alternative, in the context, is to be limited to a plane. This prevents putting elements into perspective which is possible with the use of cone trees. The radial tree is quite similar to the eye tree combining global vision of the field and the polar fish eye. But the background and focus is more significant within the graph. It appears however that the very advantages of the radial tree (focus + context) also cause a loss of contact with the primary objective, which is to keep the global view. In addition, a radial tree describing the ACM would be quite unreadable because of the huge size of the ontology. In view of the size issue, a visualization of information clusters seems to emerge through the combination of ontology and a technology called Topic Map, thanks to the open source applet Hypergraph. While not specifically conceived to effectively represent ontology, the Topic Map is a hyperbolic tree type representation, which consists of mapping the ontology and unlimited navigation. It has been adapted to enable angles of perception to stand out as well as their focus and contexts. This method will thus be a hybrid approach between the eye tree and the hyperbolic tree.

## 4 GENERATING A BROWSED META REQUEST SYSTEM

### 4.1 Meta request concept

Through the Topic Map described, OntologyNavigator provides the advanced learner access to scientific documents relating to his field of research. We intend to use external resources with OntologyNavigator, such as on-line Knowledge Base System (KBS). For this purpose we define user's context and profile to enable personal customized access to knowledge, through this application, in a transparent manner. This is a Reverse-Engineering approach of interrogation of the external KBS. A meta-query is a query sent to a remote KBS, without knowing the system of internal questioning. This is done by simulating a manual use of the remote application through combination of lemmas of keywords extracted from the context of navigation.

### 4.2 Modelling the system

While a natural language search on all words in the order established will have little chance of success, a search by key words will have every chance to return hundreds of thousands of results. The first step in generating request is the filtering of "noise" on the label when positioning the user's browser in the ontology thanks to the stop-lists (one in each language), which will eliminate empty words, like pronouns and nouns, which are too common for significant meaning. A similar preliminary stage is conducted when using a search engine in a natural language search. The second step is the lemmatization[4] of words, followed by a calculation of statistical proximity of all of the words, which have emerged, from the keyword cluster in a branch of the ontology. It may be appropriate to provide a valuation of the keywords in this context? This point could be the subject of a further study.

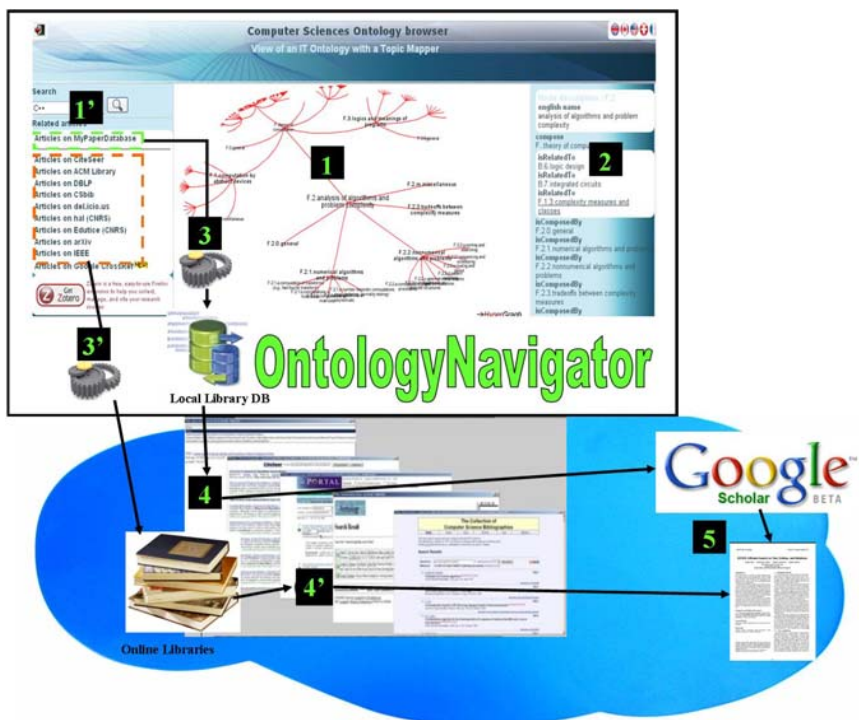

Figure 3. Functional scheme

1. and 1 ': Possibility to establish a subject's position in the ontology by browsing or through a natural language query (cf. Figure 3).

2. Positioning identifies a user point of view and interests,

---

3. and 3 ': Generating meta data and establishing requests to the RDF internally or to digital on-line libraries.

4. and 4 ': Presentation of the titles of the articles corresponding to the request and found in the RDF, or the foundation of scientific knowledge.

5. Articles are searchable on the net on Google Scholar if Uniform Resource Identifier (W3C, MIT, 2005) is absent from the base or directly available on digital libraries. If digital libraries are used, the access to a document is direct.

### 4.3 Trial of navigated search

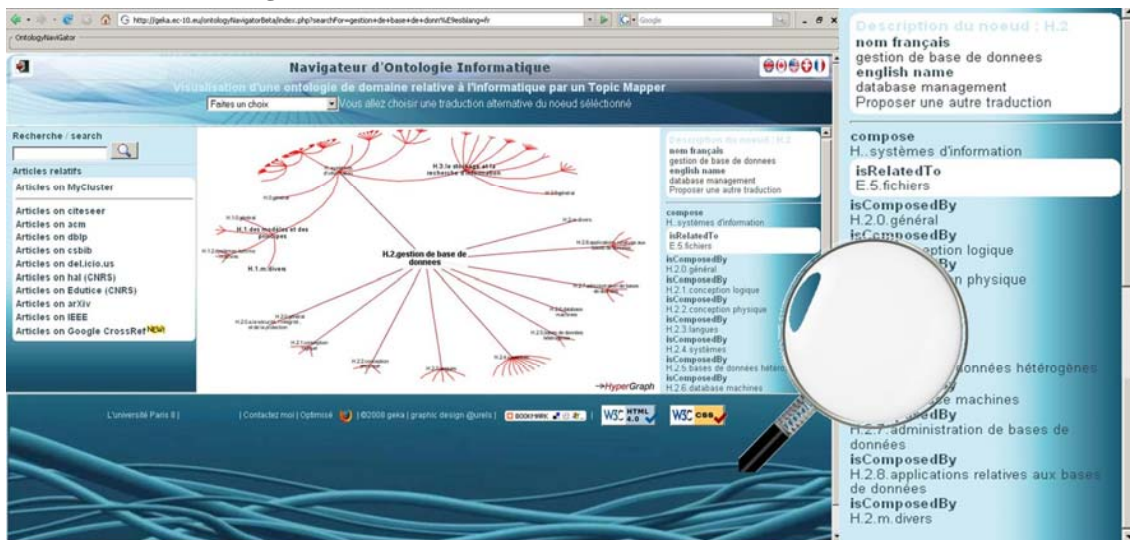

Figure 4. Search scientific papers by navigating through ontology and zoom on the focused context

The first stage of the research is to navigate down the tree until the node that is the most representative of the concept sought. The context block (cf. Figure 4) offers a direct access to on-line digital library articles as CSBIB, DBLP, or ACM by generating contextual meta-queries to these sites. In the context, queries are called meta-queries because they do not directly generate a request, but an URL with keywords. The remote Knowledge Base System (KBS) will use its own search engine to generate the real request. But the tool also proposes to search the internal database of titles of articles. In the example, search for "database management" is generated and proposes several dozen results. The article: "Managing taxonomies in relational databases" was chosen. The database provides us with the name of the principal author. The tool checks for the presence of an URI on the article in the database. In the absence of an URI a request to Google Scholar is automatically generated, which provides us with a direct access to the article (cf. Figure 5). Tests were performed on the classical databases, but results pertaining exactly to the subject of research are still too few. This mechanism for generating requests is still in a heuristic stage, but opens interesting prospects.

## 5 Results

OntologyNavigator was experimented within the University of Paris 8 (in France) in Computer Sciences department and in the Library Sciences one. It is as well accessible on-line, but the fee paid by the Library to ACM portal only grants access within the University. Nevertheless, the tool also uses several free on-line databases. A feedback form is available to get feelings and comments about the tool. The answers were used to measure final user's interest. The questions were asked about the usability, the intuitive and the results produced by the tool. Users' habits were also measured with cookies and Google Analytics. The most often comprehensive problem that users had with the tool was the lack of intuitive. The cross language search based on meta-data gave some lack of results, as described in the following paragraph.

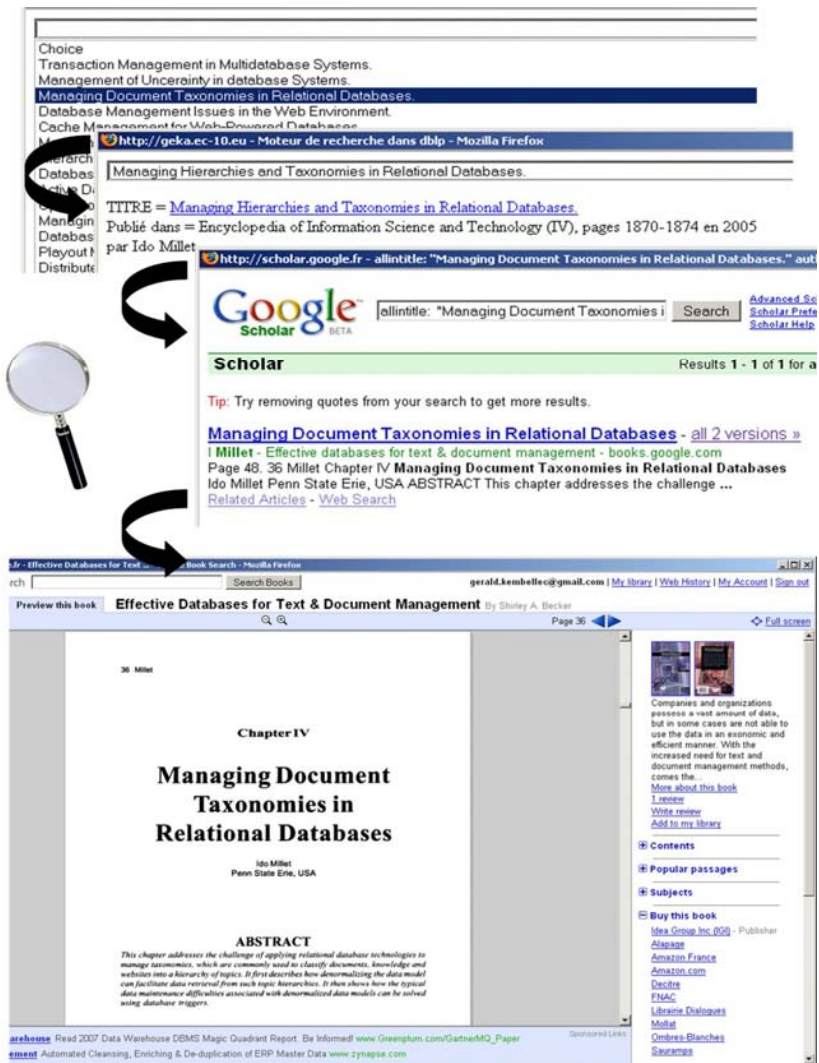

Figure 5. Scientific paper proposed by the system

A PhD candidate in computer graphics has done research in ontologyNavigator with the French sentence "rendu non photo-réaliste" (Non-Photo realistic Rendering). Research in ontology failed. The tool printed: "rendu non photo-réaliste" does not exist in French in the ACM ontology" what are known to be false in English. The articles on the NPR are usually classified under I.3 and I.4 nodes of the ACM Computer Classification System. These two nodes are respectively labelled "Computer Graphics and image processing" and "Computer vision". We then tried to use the folksonomy option for the automated processing of language to propose an alternative translation for the ontology node (cf. Figure 2). For example, the node I.3.3 "generation of images and photos" originally "picture / image generation" was given the alternative French proposal "rendu non-photoréaliste" (NPR). This proposal has no chance of being selected as the best translation by the committee of experts. This is in fact not a real translation of the node label. It is a specification and not equivalent. However, this proposal gave a result in the next time query because it created a specific entry for the French sentence "rendu non-photoréaliste". This notion, if researchers use it, allows users to include concepts of equivalence or specification outside of the simple syntax correction.

This option allows the tool to avoid terminology tendencies of the moment. For instance, in French, "rendu non-photoréaliste" (NPR) referred to the previous paragraph is not at the time of writing these lines transcription of the most widely used concept involved. The denial word "non" in "rendu non-photorealiste" conveys a negative image. Because of that fact, French specialists more likely uses "rendu expressif"

(Expressive rendering) for about two years.

The participatory community (folksonomy) also allows members to correct the shortcomings of automated processing of language. It is certain that the growing number of users of the tool significantly affect the quality of research results. This tool has the flexibility of a virtual index on a scalable corpus and presents a possible match between needs of knowledge and virtual location of on-line IT scientific articles for the young researcher.

## 6 OntologyNavigator and other KBS

| | Meta-data export available | Cross Language Research | Zotero ready | Data Mapping | Links to other KBS | Bibliographies enabled | Social Tagging |
|---|---|---|---|---|---|---|---|
| ACM Portal | yes | no | yes | no | no | no | no |
| CiteSeerX | yes | no | yes | no | yes | yes | no |
| DBLP | no | no | no | no | yes | no | no |
| ArXiv | yes | no | yes | no | no | yes | no |
| IEEE | yes | no | no | no | yes | yes | no |
| Google CrossRef | no | no | no | no | yes | no | no |
| CSBib | yes | no | yes | no | yes | no | no |
| Google scholar | no | no | no | no | yes | no | no |
| Delicious | no | no | no | no | no | yes | yes |
| HAL | no | no | yes | no | no | no | no |
| Edutice | no | no | yes | no | no | no | no |
| Ontology Navigator | yes | yes | yes | yes | yes | no | yes |

Table 1. KBS compared

Each tool offers a great number of scientific articles related to the subject sought. The idea of this comparative study is not to count the number of items returned. However, the services involved in the research were not all equal. The most often used service is the provision of meta-data in an adequate format to write a thesis or a scientific article bibliography. Usually, when such a service is available, a hyper link can generate a BibTeX or Endnote notice. The ACM portal, CiteSeer, IEEE, CSBib or OntologyNavigator, offer this service. A comparable alternative is the integration of meta-data on Article directly in XHTML thanks to micro-formats and Dublin Core. ACM, CiteSeer, CSBib, HAL, Edutice and OntologyNavigator offer this alternative service that can integrate items to Zotero. In this case, there is no need to create bibliography service within a KBS interface. CiteSeer, IEEE, ArXiv, and Delicious integrate a bibliography service.

Some databases of scientific knowledge create hyper links to other KBS from the initial research. This applies to CisteSeerX, IEEE, CSBib and DBLP. It is the main purpose of Google Scholar or OntologyNavigator. This tool dynamically creates hyper-links toward all the quoted sites in addition to its own database. Social tagging is supported natively by Delicious, because it was settled as a social tagging website. OntologyNavigator experiments social tagging and relies on researchers' involvement to enhance the quality of its translations. Moreover, it helps the community monitoring the evolution in indexed concept terminology. This tool is the only one to be able to manage to research into English language from a French request (Cross Language Retrieval).

To conclude this comparative study, OntologyNavigator provides the classical services of other KBS. Besides it adds the ability to browse through search area map navigation, which is an educational advantage. What's more OntologyNavigator offers a contextual overview that would be difficult to apprehend through a mere listing. It may happen that the articles offered by the Internal are not what the user is looking for. In this case, equivalent queries in the form of hyper-links to the main area of KBS are available. OntologyNavigator was thought to be compatible with Zotero. The two together offer the possibility to manage bibliographic records and create bibliographies. These are exported by plug-in in any format of scientific conferences, and is ready for any tool of publication (LaTeX, Word, OpenOffice...).

### 7 Limits and prospects

The testing of the tool by users has shown that the adequacy of current meta-data queries generated is relevant. Nevertheless the results are sometime poor or too big on external databases, but more precise on OntologyNavigator own database. However, the more the ontology's content is enriched with articles, the more research and indexing will become accurate. For this purpose, a pre-existing important size corpus was indexed. We will set the goal of creating a script for extracting incremental content on the on-line library DBLP updates. This ongoing automation work should refine the relevance for indexing and searching through ontology. Another limitation is the physical access to articles that is often subject to the payment of a subscription fee or even a fee per article. That is why it is easier to implement the solution in a university laboratory or a library. However, the use of proxy should help extend access to digital libraries for an entire campus. The tool will be made available to students in their second cycle of studies in the IT department and for the computation centre of the University of Paris 8. An on-line form is available for the purpose of recording feedback and follows the evolution of the users. In the near future it is planned to extend the application with an ontology based on the Friend of a Friend format (W3C, RDF and SemWeb developer, 2007) for a better understanding of the working groups, teams, and laboratories as well as links to disciplinary transversally. Another goal is to make the system as independent as possible. Possibly the hyperbolic tree / eye tree type navigation system will be modified if another way of displaying the tool emerges. To facilitate the use of items found and selected by young researchers, an interesting feature could be developed in the form of one or several thematic bibliographies on BibTeX format and thus reusable in every article and shareable with researchers with similar profiles.

### 8 Conclusion

In the French University context, advanced learners, in 2 and 3 study cycles (end of graduate and postgraduate studies) often experiencing real difficulties in collecting the documentation in their field of study or research. The main purpose is to help these learners by providing a tool to supplement their perception of the knowledge domain "stored" in ontology. It is to consider, even hoping, that the tool will become obsolete by his mastery (because mastery of the tool is an intrinsic source of knowledge). This study focuses on bibliographical research in the IT field by "insiders" but not experts, i.e. young researchers that are increasingly lost in a predominantly English corpus. The basis of this research is a reflection on the concept of projecting domain ontology on the concepts of portal and search engine relevance. This approach tends to generate a transparent and intuitive man-machine interface (MMI) in order to improve the user approach. The question raised by this discussion is the impact of knowledge representation on information retrieval and learning. Is the impact of the tool on the results obtained as compared to more traditional research significant? The ultimate goal of this research is to develop an incremental system and ideally an autonomous system capable of indexing scientific documents by extracting keywords and placing articles in IT domain ontology. The autonomy of the system would be a significant factor in cost reduction. More importantly it could avoid having a group of experts lost in an endless work with every technical or ideological detail which would provoke discussions on the appropriateness of indexing a new concept or not, and on the place of the new concept on the ontology. This approach will also provide easier access to information sought by search engines using natural language, keywords, context, or semantic proximity. Thus, and this is the key concept, even a user who does not master the entire computer vocabulary (and the English language) might find relevant articles in several languages which he would not have accessed through traditional research methods. This article proposes to implement the first part of this task, namely the construction of ontology, and of a navigation system to browse and search the ontology in a scientific corpus. In this work, a research tool for scholars whose work is related to IT as been created. This consists in an on-line interface to link an ontology search to on-line scientific libraries. The ontology-based ACM CCS has been translated into French in an automatic way to assist researchers wishing to conduct research in French. This solution offers researchers a possibility to find articles on a study revolving around a node in the IT ontology field. This query is generated by graphic navigation or natural language queries. Once the research is completed an automated system can find articles on the internal database or propose meta-data queries to scientific on-line digital libraries. The system generation of meta-search is based solely on the titles of nodes in the ontology. However, the results are encouraging and the prospects for improvement in the near future are already under consideration. For the time being, OntologyNavigator works only on the IT field. Very advanced domain ontologies can also be found in biomedical, standards and safety of the building construction (Barzic, J. 2008), and law fields. The application can be exported in any of these areas by changing the KBS and ontology domains.